\documentclass[a4paper]{article}
\usepackage[german, english]{babel}
\usepackage{a4wide}
\usepackage{amsmath}
\usepackage{amssymb}
\usepackage{ifthen}
\usepackage{epsfig}
\newcounter{fig}    

\newcommand{\vphi}{\varphi}
\newcommand{\Tr}{{\rm Tr}}

\begin{document}

\vspace*{2cm}

\begin{center}
{\bf\Large Monopole--Antimonopole Chains}
\vspace{1.0cm}

{\sc\bf  Burkhard Kleihaus}\\[10pt]
{\it Department of Mathematical Physics, University College, Dublin,\\
Belfield, Dublin 4, Ireland}\\[12pt]
{\sc\bf  Jutta Kunz} and {\sc\bf Yasha Shnir}\\[10pt]
{\it Institut f\"ur Physik, Universit\"at Oldenburg}\\
{\it D-26111, Oldenburg, Germany}

\end{center}

%\maketitle 

%\thispagestyle{empty}

\bigskip

%-----------------------------------
%   Abstract
%-----------------------------------
\begin{abstract}
We present new static axially symmetric solutions of SU(2) 
Yang-Mills-Higgs theory, representing chains of monopoles
and antimonopoles in static equilibrium.
They correspond to
saddlepoints of the energy functional and exist both in the 
topologically trivial sector and in the sector with topological charge one.   
\end{abstract}

\section{Introduction }
 
Magnetic monopole solutions are a generic prediction of grand unified theories.
Such solutions possess a topological charge proportional to 
their magnetic charge.
In Yang-Mills-Higgs (YMH) theory with gauge group SU(2)
the simplest solution with unit topological charge is the
spherically symmetric 't~Hooft-Polyakov
monopole \cite{Hooft74,Polyakov74}.
SU(2) monopoles with higher topological charge cannot 
be spherically symmetric \cite{WeinbergGuth}
and possess at most axial symmetry \cite{RebbiRossi,Ward,Forgacs,Prasad}.

In the Bogomol'nyi-Prasad-Sommerfield (BPS) limit of vanishing Higgs potential
spherically symmetric monopole and axially symmetric multimonopole solutions, 
which satisfy the first order Bogomol'nyi equations \cite{Bogo}
as well as the second order field equations,
are known analytically \cite{PraSom,Ward,Forgacs,Prasad}. 
For these solutions all zeros of the Higgs field are 
superimposed at a single point.  
Multimonopole solutions of the Bogomol'nyi equations
which do not possess any rotational symmetry \cite{CorGod},
have recently been constructed numerically
\cite{monoDS}. In these solutions
the zeros of the Higgs field are no longer all
superimposed at a single point but are located at
several isolated points.

As shown by Taubes, in each topological sector 
there exist in addition smooth, finite energy solutions of the second order
field equations, which do not satisfy the Bogomol'nyi equations
\cite{Taubes85}. Consequently, the energy of these solutions exceeds
the Bogomol'nyi bound.
The simplest such solution resides in the topologically 
trivial sector and forms a saddlepoint of the energy functional
\cite{Taubes82}.   
It possesses axial symmetry, and the two zeros of its
Higgs field are located symmetrically on the positive and negative 
$z$-axis. 
This solution corresponds to a monopole and 
antimonopole in static equilibrium \cite{Rueber,mapKK}.

In this paper we present new axially symmetric 
saddlepoint solutions,
where the Higgs field vanishes at $m>2$ isolated points on
the symmetry axis. 
These configurations represent 
chains of monopoles and antimonopoles, in alternating order.
Chains with an equal number of monopoles and antimonopoles, $m=2k$,
reside in the topologically trivial sector,
whereas chains with $k+1$ monopoles and $k$ antimonopoles
reside in the sector with topological charge one.

After briefly reviewing  SU(2) Yang-Mills-Higgs theory in Section 2,
we discuss the axially symmetric Ansatz in Section 3
together with the boundary conditions. 
We present the numerical results in Section 4 and our conclusions
in Section 5.

\section{Yang-Mills-Higgs theory}

The Lagrangian of SU(2) Yang-Mills-Higgs theory is given by
\begin{equation}
-L = \int\left\{ \frac{1}{2} \Tr\left( F_{\mu\nu} F^{\mu\nu}\right)
                +\frac{1}{4} \Tr\left( D_\mu \Phi D^\mu \Phi \right)
		+\frac{\lambda}{8} \Tr\left[ \left(\Phi^2 - \eta^2\right)^2
 \right]
	\right\} d^3 r \ ,
\label{lag}
\end{equation}
with su(2) gauge potential $A_\mu = A_\mu^a \tau^a/2$,
field strength tensor
$F_{\mu\nu} = \partial_\mu A_\nu - \partial_\nu A_\mu + i e [A_\mu, A_\nu]$,
and covariant derivative of the Higgs field
$D_\mu \Phi = \partial_\mu \Phi +i e [A_\mu, \Phi]$.
$e$ denotes the gauge coupling constant, $\eta$ the vacuum expectation
value of the Higgs field and $\lambda$ the strength of the Higgs selfcoupling.

The field equations are derived from the Lagrangian
as the variational equations with respect to the gauge potential and the
Higgs field,
\begin{eqnarray}
D_\mu F^{\mu\nu} 
-\frac{1}{4}i e \left[ \Phi, D^\nu \Phi\right]
& = & 0 \ ,
\label{Feq} \\
D_\mu  D^\mu \Phi 
- \lambda\left(\Phi^2-\eta^2\right) \Phi
& = & 0 \ .
\label{Higgseq}
\end{eqnarray}

\section{Ansatz and Boundary Conditions}

We parametrize the gauge potential and the Higgs field by the Ansatz
\begin{eqnarray}
A_\mu dx^\mu
& = &
\left( \frac{K_1}{r} dr + (1-K_2)d\theta\right)\frac{\tau_\vphi}{2e}
- \sin\theta \left( K_3\frac{\tau_r^{(m)}}{2e}
                     +(1-K_4)\frac{\tau_\theta^{(m)}}{2e}\right) d\vphi
\label{ansatzA} \\
\Phi
& = &
\Phi_1\tau_r^{(m)}+ \Phi_2\tau_\theta^{(m)} \  .
\label{ansatzPhi}
\end{eqnarray}
which generalizes the axially symmetric Ansatz employed in 
\cite{Rueber,mapKK} 
for the monopole-antimonopole pair solution.
Here the su(2) matrices
$\tau_r^{(m)}$, $\tau_\theta^{(m)}$, and $\tau_\vphi$
are defined as
\begin{eqnarray}
\tau_r^{(m)}  & = &
\sin(m\theta) \tau_\rho + \cos(m\theta) \tau_z \ ,
\nonumber\\
\tau_\theta^{(m)} & = &
\cos(m\theta) \tau_\rho - \sin(m\theta) \tau_z \ ,
\nonumber\\
\tau_\vphi & = &
 -\sin \vphi \tau_x + \cos \vphi\tau_y \ ,
\nonumber
\end{eqnarray}
where $\tau_\rho = \cos \vphi \tau_x + \sin \vphi\tau_y $,
and $m$ is an integer to which we refer as $\theta$ winding number.
 
The profile functions $K_1$ -- $K_4$ and $\Phi_1$, $\Phi_2$ 
depend on the coordinates $r$
and $\theta$, only. With this Ansatz the general field equations Eqs.~(\ref{Feq}) and
(\ref{Higgseq}) reduce to six PDEs in the coordinates $r$ and $\theta$.
The ansatz possesses a residual U(1) gauge symmetry. To fix the gauge we impose
the condition $r\partial_r K_1 - \partial_\theta K_2 = 0$. For
convenience we change to dimensionless coordinates by rescaling 
$r\to r/(e\eta)$ and $\Phi \to \eta \Phi$. 

To obtain regular solutions with finite energy and energy density we have to 
impose appropriate boundary conditions. 
Regularity at the origin requires
$$
K_1(0,\theta)=0\ , \ \ \ \ K_2(0,\theta)= 1 \ , \ \ \ \
K_3(0,\theta)=0 \ , \ \ \ \ K_4(0,\theta)=1 \ , \ \ \ \
$$
$$
\sin(m\theta) \Phi_1(0,\theta) + \cos(m\theta) \Phi_2(0,\theta) = 0 \ ,
$$
$$
\left.\partial_r\left[\cos(m\theta) \Phi_1(r,\theta)
              - \sin(m\theta) \Phi_2(r,\theta)\right] \right|_{r=0} = 0 .
$$
To obtain the boundary conditions at infinity we require that
solutions in the vacuum sector ($m=2k$) tend to
a gauge transformed trivial solution, 
$$
\Phi \ \longrightarrow U \tau_z U^\dagger \   , \ \ \
A_\mu \ \longrightarrow  \ i \partial_\mu U U^\dagger \ ,
$$
and that solutions in the sector with unit topological charge ($m=2k+1$)
tend to
$$
\Phi  \longrightarrow  U \Phi_\infty^{(1)} U^\dagger \   , \ \ \
A_\mu \ \longrightarrow \ U A_{\mu \infty}^{(1)} U^\dagger
+i \partial_\mu U U^\dagger \  ,
$$
where
$$ \Phi_\infty^{(1)} =\tau_r^{(1)}\ , \ \ \
A_{\mu \infty}^{(1)}dx^\mu =
\frac{\tau_\vphi}{2} d\theta
- \sin\theta \frac{\tau_\theta^{(1)}}{2} d\vphi
$$
is the asymptotic solution of a monopole,
and  $U = \exp\{-i k \theta\tau_\vphi\}$, both
for even and odd $m$.
Consequently, solutions with even $m$ have vanishing magnetic charge,
whereas solutions with odd $m$ possess unit magnetic charge.

In terms of the functions $K_1 - K_4$, $\Phi_1$ and $\Phi_2$ these boundary
conditions read
\begin{equation}
K_1 \longrightarrow 0 \ , \ \ \ \
K_2 \longrightarrow 1 - m \ , \ \ \ \
\label{K12infty}
\end{equation}
\begin{equation}
K_3 \longrightarrow \frac{\cos\theta - \cos(m\theta)}{\sin\theta}
\ \ \ m \ {\rm odd} \ , \ \ \
K_3 \longrightarrow \frac{1 - \cos(m\theta)}{\sin\theta}
\ \ \ m \ {\rm even} \ , \ \ \
\label{K3infty}
\end{equation}
\begin{equation}
K_4 \longrightarrow 1- \frac{\sin(m\theta)}{\sin\theta} \ ,
\label{K4infty}
\end{equation}
\begin{equation}
\Phi_1\longrightarrow  1 \ , \ \ \ \ \Phi_2 \longrightarrow 0 \ .
\label{Phiinfty}
\end{equation}
Regularity on the $z$-axis requires
$$
K_1 = K_3 = \Phi_2 =0 \ , \ \ \  \
\partial_\theta K_2 = \partial_\theta K_4 = \partial_\theta \Phi_1 =0 \ ,
$$
for $\theta = 0$ and $\theta = \pi$.

\section{Results}

We have constructed numerically axially symmetric solutions 
of the Yang-Mills-Higgs theory in the BPS limit $\lambda = 0$ for
$\theta$ winding number $1\leq m \leq 6$.

These $m$-chains possess $m$ nodes of the Higgs field on the $z$-axis.
Due to reflection symmetry, each node on the negative $z$-axis corresponds
to a node on the positive $z$-axis.
The nodes of the Higgs field are associated with the location of the
monopoles and antimonopoles. 
For odd $m$ ($m=2k+1$) the Higgs field possesses 
$k$ nodes on the positive $z$-axis and one node at the origin.
The node at the origin corresponds
to a monopole if $k$ is even and to an antimonopole if $k$ is odd.
For even $m$ ($m=2k$) there is 
no node of the Higgs field at the origin. 

The $m=1$ solution is the well-known 't Hooft-Polyakov monopole
\cite{Hooft74,Polyakov74}. 
The $m=3$ (M-A-M) and $m=5$ (M-A-M-A-M) chains represent
saddlepoints with unit topological charge.
The $m=2$ chain is identical to the 
monopole-antimonopole (M-A) pair discussed in \cite{Rueber,mapKK}.
The M-A pair as well as the $m=4$ (M-A-M-A) and $m=6$ (M-A-M-A-M-A)
chains form saddlepoints in the vacuum sector. 

In Table 1 we present the dimensionless energy $E/4\pi\eta$, 
the dimensionless magnetic dipole moment $\mu$ and the nodes $z_i$ of 
the Higgs field for the solutions with 
$\theta$ winding number $1\leq m \leq 6$.

We observe that the energy  $E^{(m)}$ of an $m$-chain
is always smaller than the energy 
of $m$ single monopoles or antimonopoles (with infinite separation
between them),
i.~e.~$E^{(m)} < E_\infty /4\pi\eta = m$. 
On the other hand $E^{(m)}$ exceeds the minimal energy bound
given by the Bogolmol'ny limit $E_{\rm min}/4\pi\eta = 0$ for even $m$,
and $E_{\rm min}/4\pi\eta = 1$ for odd $m$.
We observe an (almost) linear dependence of the energy $E^{(m)}$ on $m$.
This can be modelled
by taking into account only the energy of $m$ single
(infinitely separated)
monopoles and the next-neighbour interaction between monopoles
and antimonopoles on the chain.
Defining the interaction energy as the binding energy of the 
monopole-antimonopole pair, 
$$
\Delta E = 2\ (4\pi \eta) - E^{(2)} \ ,
$$
we obtain as energy estimate for the $m$-chain
$$
E_{\rm est}^{(m)}/4\pi \eta  = m +(m-1) \Delta E \ .
$$ 
In Fig.~1 we show this estimate for the energy, $E_{\rm est}^{(m)}/4\pi \eta$,
together with the exact energy for chains with $\theta$ winding number 
$m=1, \dots , 6$.
We note that the deviation of the estimated energy from the exact energy is 
indeed very small.

In Table 2 we list the possible decompositions of the $m$-chains 
into subchains of smaller length together with the total energy of the 
subchains (at infinite separation). We observe that the energetically 
most favourable state is the $m$-chain. This supports our
interpretation of the $m$-chain as an equilibrium state of 
$m$ monopoles and antimonopoles. We note, however, that there are 
other possible decompositions not included in the table.
For example, the M-A-M-A-chain could be decomposed into
a pair of a charge $2$ multimonopole and a charge $-2$ anti-multimonopole.

To define the magnetic dipole moment for solutions with even $m$,
we first transform to a gauge where the Higgs field
is constant at infinity,
$\Phi = \tau_z$. From the asymptotic expansion \cite{long}, we obtain
$K_3 \to (1-\cos(m \theta))/\sin(\theta) +C_3 \sin\theta/r$.
Thus the gauge field assumes the asymptotic form 
\begin{equation}
A_\mu dx^\mu = {C_3}\frac{\sin^2\theta}{2r}\tau_z d\vphi \ , 
\label{magmom}
\end{equation}
from which we read off the (dimensionless) magnetic dipole moment 
$\mu= C_3$.
Solutions with odd $m$ have vanishing magnetic dipole moment,
since in this case the function $K_3$ is odd under the transformation
$z \leftrightarrow -z$. Consequently, the asymptotic form of the 
gauge potential cannot contain terms like the r.h.s.~of Eq.~(\ref{magmom}).

The magnetic dipole moment also increases with increasing $m$,
as seen in Table 1. 
To obtain an estimate for the dipole moment we consider the magnetic 
charges as point charges located at the nodes of the Higgs field,
yielding
$$
\mu_{\rm est}^{(m)} = \sum_{i=1,m} z_i P_i , 
$$
with charges $P_i = 1$ for monopoles and  $P_i =-1$ for antimonopoles, 
respectively.
For comparison the estimated magnetic dipole moments are also given 
in Table 1.  The deviation from the exact values is 
on the order of $10$\%.

Concerning the nodes of the Higgs field
we observe that the distances between the nodes increase with 
increasing $m$. 
Remarkably, the distances between the nodes do 
not vary much within a chain. For example, denoting the 
location of the nodes by $z_i$ in decreasing order we find for 
the chain with $\theta$ winding number $m=6$
from Table 1,
$|z_1-z_2| \approx 5.06$, $|z_2-z_3| \approx 5.11$ and 
$|z_3-z_4| \approx 4.92$.

In Fig.~2 we present the dimensionless energy density
for the solutions with $\theta$ winding number $m=1, \dots , 6$. 
The energy density of the $m$-chain possesses $m$ maxima on the $z$-axis,
and decreases with increasing $\rho$.
The locations of the maxima are close to the nodes of the Higgs field,
which are indicated by asterisks. For a given $m$ the maxima are of 
similar magnitude, but their height decreases with increasing $m$.
(Note that the scale for the $m=1$ solution is different compared to 
the $m\geq 2$ solutions, and that 
the contour lines are distorted due to the different scaling of 
the $\rho$- and $z$-axis.)

\section{Conclusion}

We have obtained new static axially symmetric solutions of the 
SU(2) Yang-Mills-Higgs theory which represent 
monopole-antimonopole chains. They are characterized by the 
$\theta$ winding number $m$, which equals the number of nodes of the 
Higgs field, and the total number of monopoles and antimonopoles.
Solutions with even $m$ carry no magnetic charge but possess a non-vanishing 
magnetic dipole moment, whereas
solutions with odd $m$ carry unit magnetic charge but possess 
no magnetic dipole moment.
The energy of these $m$-chains increases (almost) linearly with $m$.

We interpret the $m$-chains as equilibrium states of $m$
monopoles and antimonopoles. As shown long ago \cite{Manton77},
the force between monopoles is given by twice the Coloumb force
when the charges are unequal, and vanishes when the charges are equal,
provided the monopoles are at large distances.
Thus, monopoles and antimonopoles can only be 
in static equilibrium, if they are close enough to experience a
repulsive force that counteracts the attractive Coloumb force.
In other words, $m$-chains are essentially non-BPS solutions.
To see this in another way let us rewrite the energy in the 
form
\begin{equation}
E  = \int\left\{ \frac{1}{4} \Tr\left(
\left(\varepsilon_{ijk} F_{ij}\pm D_k \Phi \right)^2 \right)
 \mp\frac{1}{2}\varepsilon_{ijk} \Tr\left( F_{ij} D_k \Phi \right)	
	\right\} d^3 r \ .
\label{E2}
\end{equation}
The second term is proportional to the topological charge and 
vanishes when $m$ is even. The first term is just the integral
of the square of the Bogol'molnyi equations. Thus, for even $m$ 
the energy is a measure for the deviation of the solution from 
selfduality.

So far we have considered only solutions in the vacuum sector
or the sector with unit topological charge. Generalising the ansatz
to $\vphi$ winding number $n>1$ leads to new solutions, 
which carry topological charge $n$ if $m$ is odd, and to solutions 
in the vacuum sector if $m$ is even \cite{KKS-2}.
Recently, the $n=2$, $m=2$ solution has been constructed 
in an extended model \cite{Tigran}, which includes the YMH theory as a 
special case.
Chains of $n=2$ multimonopoles and anti-multimonopoles will 
be presented elsewhere \cite{KKS-2}.
For $n>2$ a new phenomenon occurs. The zeros of the Higgs field
no longer form a set of isolated points. Instead the Higgs field 
vanishes on rings around the $z$-axis \cite{KKS-2}. (For odd $m$ the node at 
the origin persists). 

Dyonic solutions can be readily obtained from the 
$m$-chains \cite{BijRadu,long}, as outlined in \cite{HKK,WeinbergDy}.
Interestingly, these solutions carry electric charge even for solutions
in the vacuum sector.

When the gravitional interaction is included, we anticipate a different 
behaviour for $m$-chains with finite  magnetic charge and those with
vanishing magnetic charge.
For magnetically charged solutions a degenerate horizon may form
at a critical value of the gravitional parameter, as observed for 
monopoles \cite{gravMP} and multimonopoles \cite{gravMMP}.  
On the other hand, no formation of a horizon was found 
for the gravitating monopole-antimonopole pair \cite{gravMAP}.

We expect that solutions analogous to the $m$-chains exists in 
Weinberg-Salam theory \cite{Klinkhamer,Sstar,BrihayeKunz},
generalizing the sphaleron-antisphaleron pair \cite{Sstar}.
The axially symmetric Ansatz with $\vphi$ winding number $n$ 
and $\theta$ winding number $m$ \cite{BrihayeKunz} allows for 
(multi)sphaleron--anti-(multi)sphaleron chains and 
solutions with rings of zeros.

\newpage
%%%%%%%%%%%%%%%%%%%%%%%%%%%%%%%%%%%%%%%%%%%%%%%%%%%%%%%%%%%%%%%%%%%%%%%%%
 \begin{center}
\begin{tabular}{|c|c|c|c|c|}
 \hline
 m  &  $E[4\pi\eta]$   & $\mu$   & $\mu_{\rm est}$   & $z_i$\\
 \hline
1     & $1.00$  & $0.0  $ & $0.0  $ & $0.0$ \\
2     & $1.70$  & $4.72 $ & $4.18 $ & $\pm 2.09$ \\
3     & $2.44$  & $0.0  $ & $0.0  $ & $0.0 $ $\pm 4.67$ \\
4     & $3.12$  & $9.86 $ & $9.21 $ & $\pm 2.39 \ $ $\pm 6.99$ \\
5     & $3.78$  & $0.0  $ & $0.0  $ & $0.0 \ $ $\pm 4.79 \ $ $\pm 9.61$ \\
6     & $4.40$  & $16.06$ & $15.40$ & $\pm 2.46 \ $ $\pm 7.57 \ $ $\pm 12.63$ \\
 \hline
\end{tabular}\vspace{7.mm}\\
{\bf Table 1}
The dimensionless energy, the dipole moment $\mu$, 
the estimated dipole moment $\mu_{\rm est}^{(m)}$, and 
the nodes $z_i$ are given for the $m$-chains with $m=1,\dots ,6$.
\vspace{7.mm}\\
\end{center}%\end{table}
 \begin{center}
\begin{tabular}{|c|c|c|c|c|}
% \hline
\multicolumn{3}{c}{} & \multicolumn{2}{c}{} \\ 
 \hline
m& chain  &  $E[4\pi\eta]$ &  decomposition   & $E[4\pi\eta]$\\
 \hline
2 & M-A     & $E^{(2)}=1.70$  &  (M)+(A)  & $2 E^{(1)}= 2.00 $ \\
 \hline
3 & M-A-M    & $E^{(3)}=2.44$  &
\begin{tabular}{c}
 (M-A)+(M)  \\ 
 (M)+(A)+(M)
\end{tabular}
&
\begin{tabular}{c}
$ E^{(2)}+E^{(1)}=2.7$ \\
$ 3 E^{(1)}=3.00$
\end{tabular}
\\
 \hline
4 & M-A-M-A   & $E^{(4)}=3.12$  & 
\begin{tabular}{c}
(M-A)+(M-A)\\ 
(M-A-M)+(A)\\
(M-A)+(M)+(A)\\
(M)+(A)+(M)+(A)
\end{tabular}
&
\begin{tabular}{c}
$ 2E^{(2)}=3.40$\\
$ E^{(3)}+E^{(1)}=3.44$\\
$ E^{(2)}+2E^{(1)}=3.70$\\
$ 4E^{(1)}=4.00$
\end{tabular}\\
 \hline
5 & M-A-M-A-M  & $E^{(5)}=3.78$  & 
\begin{tabular}{c}
(M-A-M-A)+(M)\\
(M-A-M)+(A-M)\\
(M-A)+(M-A)+(M)\\
(M-A-M)+(A)+(M)\\
(M-A)+(M)+(A)+(M)\\
(M)+(A)+(M)+(A)+(M)
\end{tabular}
&
\begin{tabular}{c}
$ E^{(4)}+E^{(1)}=4.12$\\
$ E^{(3)}+E^{(2)}=4.14$\\
$ 2E^{(2)}+E^{(1)}=4.40$\\
$ E^{(3)}+2E^{(1)}=4.44$\\
$ E^{(2)}+3E^{(1)}=4.7$\\
$ 5E^{(1)}=5.00$
\end{tabular}\\
 \hline
6 & M-A-M-A-M-A & $E^{(6)}=4.40$  & 
\begin{tabular}{c}
(M-A-M-A-M)+(A)\\
(M-A-M-A)+(M-A)\\
(M-A-M)+(A-M-A)\\
(M-A)+(M-A)+(M-A)\\
(M-A-M-A)+(M)+(A)\\
(M-A-M)+(A-M)+(A)\\
(M-A)+(M-A)+(M)+(A)\\
(M-A-M)+(A)+(M)+(A)\\
(M-A)+(M)+(A)+(M)+(A)\\
(M)+(A)+(M)+(A)+(M)+(A)
\end{tabular}
&
\begin{tabular}{c}
$ E^{(5)}+E^{(1)}=4.78$\\
$ E^{(4)}+E^{(2)}=4.82$\\
$ 2E^{(3)}=4.88$\\
$ 3E^{(2)}=5.10$\\
$ E^{(4)}+2E^{(1)}=5.12$\\
$ E^{(3)}+E^{(2)}+E^{(1)}=5.14$\\
$ 2E^{(2)}+2E^{(1)}=5.40$\\
$ E^{(3)}+3E^{(1)}=5.44$\\
$ E^{(2)}+4E^{(1)}=5.70$\\
$ 6E^{(1)}=6.00$
\end{tabular}\\
 \hline
\end{tabular}\vspace{7.mm}\\
{\bf Table 2}
The decompositions of the $m$-chains into subchains and their 
energies are given for $m=1,\dots ,6$.
\vspace{7.mm}\\
\end{center}%\end{table}
%%%%%%%%%%%%%%%%%%%%%%%%%%%%%%%%%%%%%%%%%%%%%%%%%%%%%%%%%%%%%%%%%%%%%%%%%
% fig 1
%\vspace{7mm}\\  
\parbox{\textwidth}
{\centerline{
\mbox{
\epsfysize=12.0cm
\epsffile{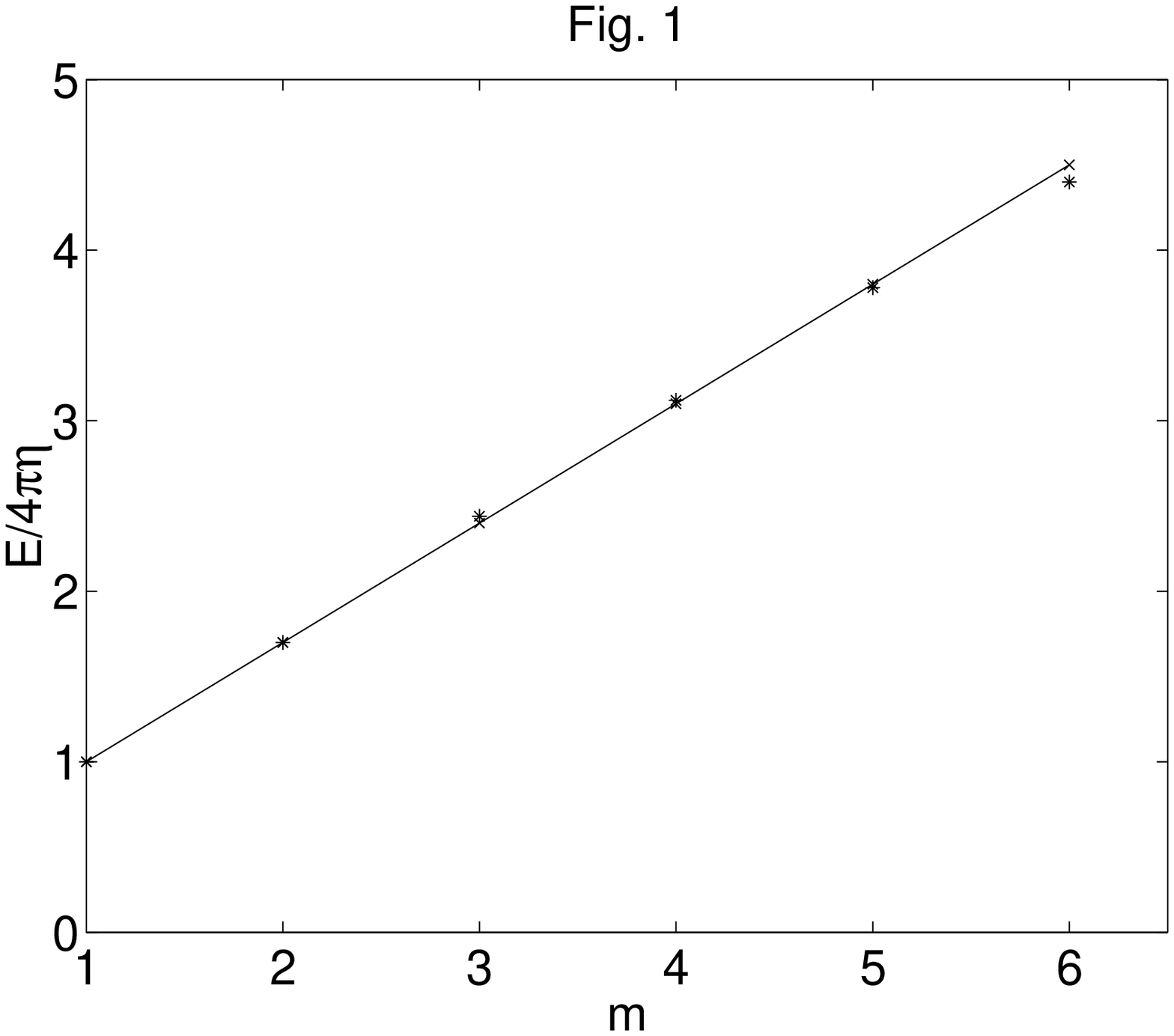}
}
}
}
{\bf Fig.~1} The exact energy (asterisks) and the estimate for the
 energy (crosses)  are shown for chains with $m=1,\dots ,6$. 
The solid line demonstrates the linear dependence of 
$E_{\rm est}^{(m)}$ on $m$.
\newpage
%%%%%%%%%%%%%%%%%%%%%%%%%%%%%%%%%%%%%%%%%%%%%%%%%%%%%%%%%%%%%%%%%%%%
% fig 2
\vspace*{-4.cm}
\parbox{\textwidth}
{
\centerline{
\mbox{
\epsfysize=29.0cm
\epsffile{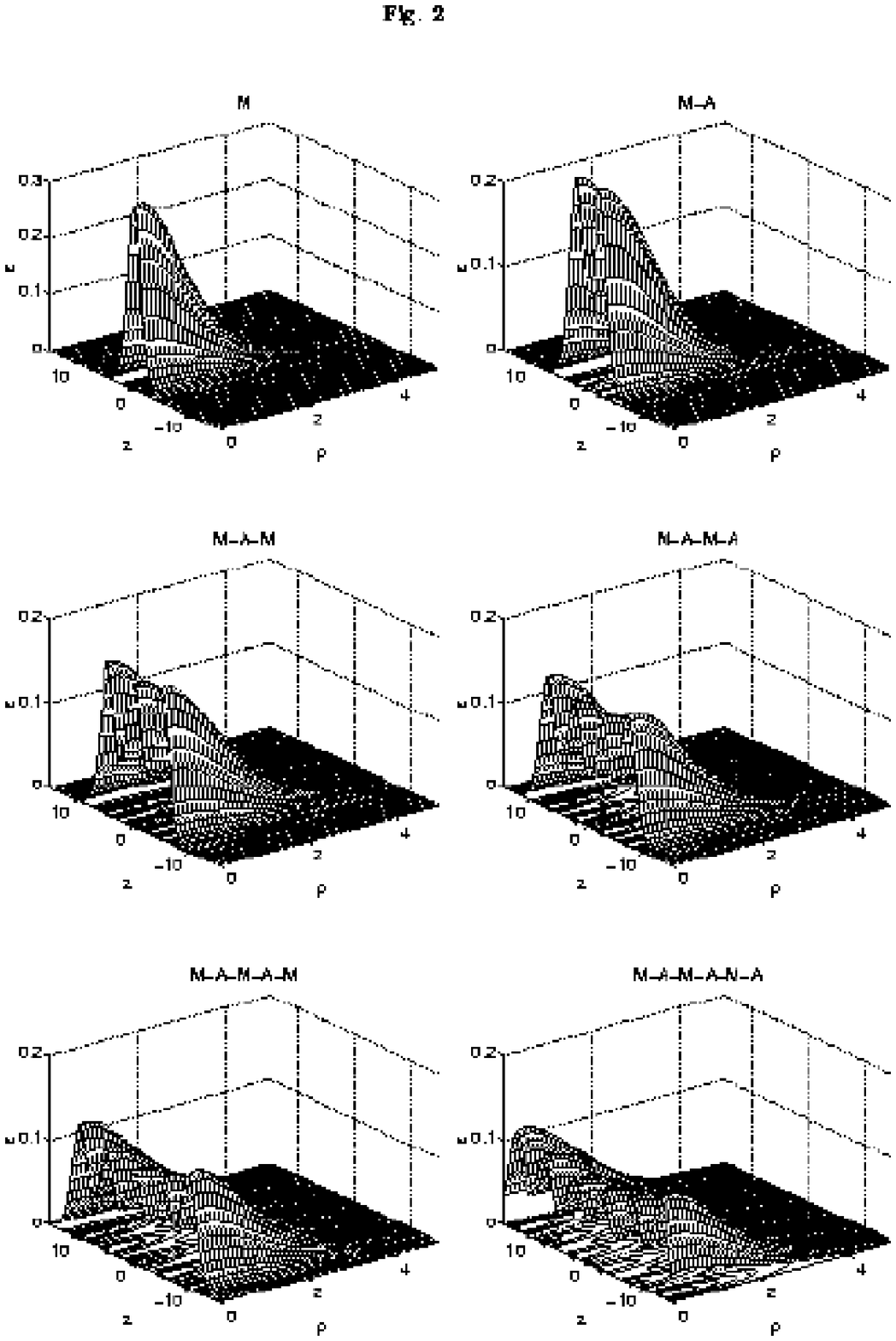}
}
}
}\vspace*{-4.cm}
{\bf Fig.~2} The dimensionless energy density is shown as 
function of $\rho$ and $z$ for chains with $m=1, \dots ,6$.
The asterisks indicate the nodes of the Higgs field.
%%%%%%%%%%%%%%%%%%%%%%%%%%%%%%%%%%%%%%%%%%%%%%%%%%%%%%%%%%%%%%%%%%%%

\begin{thebibliography}{000}
\bibitem{Hooft74}G.~`t Hooft, Nucl.\ Phys.\ {\bf B79} (1974) 276.

\bibitem{Polyakov74}A.M.Polyakov, Pis'ma JETP  {\bf 20} (1974)  430.

\bibitem{WeinbergGuth}E.J.~Weinberg and A.H.~Guth, 
                      Phys. Rev. {\bf D14} (1976) 1660.

\bibitem{RebbiRossi}C.~Rebbi and P.~Rossi,  Phys. Rev. {\bf D22} (1980) 2010.
 
\bibitem{Ward}R.S.~Ward, Comm. Math. Phys., {\bf 79} (1981) 317.

\bibitem{Forgacs}P.~Forgacs, Z.~Horvath and L.~Palla,
                 Phys. Lett., {\bf 99B} (1981) 232.

\bibitem{Prasad}M.K.~Prasad,  Comm. Math. Phys., {\bf 80} (1981) 137;
                M.K.~Prasad and P.~Rossi,   Phys. Rev. {\bf D24} (1981) 2182.

\bibitem{Bogo}E.B. Bogomol'nyi,
              Yad.~Fiz. {\bf 24} (1976) 861 
              [Sov.~J.~Nucl.~Phys. {\bf 24}, (1976) 449]. 

\bibitem{PraSom}M.K.~Prasad and C.M.~Sommerfeld, 
                Phys. Rev. Lett. {\bf 35} (1975) 760.

\bibitem{CorGod} E.~Corrigan and P. Goddard, 
                 Comm. Math. Phys., {\bf 80} (1981) 575.

\bibitem{monoDS}
see e.g.~P.~M. Sutcliffe, 
         Int. J. Mod. Phys. {\bf A 12} (1997) 4663;
         C.~J. Houghton, N.~S. Manton and P.~M. Sutcliffe, 
         Nucl. Phys. {\bf B 510} (1998) 507.

\bibitem{Taubes85}C.~H. Taubes,
                  Commun. Math. Phys. {\bf 97} (1985) 473 

\bibitem{Taubes82}C.~H. Taubes,
                  %The existence of a non-minimal solution
                  %to the SU(2) Yang-Mills-Higgs equations on $R^3$. Part I,
                  Commun. Math. Phys. {\bf 86} (1982) 257; 
                  %Part II,
                  ibid 
                  %Commun. Math. Phys. 
                  {\bf 86} (1982) 299.

\bibitem{Rueber} Bernhard R\"uber,
                 %Eine axialsymmetrische magnetische Dipoll\"osung der
                 %Yang-Mills-Higgs-Gleichungen,
                 Thesis, University of Bonn 1985.

\bibitem{mapKK} B.~Kleihaus and J.~Kunz,
                %A monopole antimonopole solution of the SU(2) Yang-Mills-Higgs
                % model,
                 Phys. Rev. {\bf D61} (2000) 025003.

\bibitem{long} B.~Kleihaus, J.~Kunz and Ya.~Shnir, in preparation.

\bibitem{Manton77} N.S.~Manton, 
                   Nucl. Phys. {\bf B 126} (1977) 525.

\bibitem{KKS-2} B.~Kleihaus, J.~Kunz and Ya.~Shnir, in preparation.

\bibitem{Tigran}  V. Paturyan and D.H.~Tchrakian,
                  Monopole Antimonopole solutions of the skyrmed 
                  SU(2) Yang-Mills-Higgs model, hep-th/0306160.

\bibitem{BijRadu}J.~J.~Van der Bij and E.~Radu,
                 Int.\ J.\ Mod.\ Phys.\ A {\bf 17} (2002) 1477.

\bibitem{HKK} B.~Hartmann, B.~Kleihaus and J.~Kunz,
              Mod.\ Phys.\ Lett.\ A {\bf 15} (2000) 1003.

\bibitem{WeinbergDy}E.~J.~Weinberg,
                    Phys.\ Rev.\ D {\bf 20} (1979) 936.

\bibitem{gravMP}P.~Breitenlohner, P.~Forgacs and D.~Maison,
                Nucl.\ Phys.\ B {\bf 383} (1992) 357;
                Nucl.\ Phys.\ B {\bf 442} (1995) 126.

\bibitem{gravMMP} B.~Hartmann, B.~Kleihaus and J.~Kunz,
                  Phys.\ Rev.\ Lett.\  {\bf 86} (2001) 1422;
                  Phys.\ Rev.\ D {\bf 65} (2002) 024027. 

\bibitem{gravMAP} B.~Kleihaus and J.~Kunz,
                  Phys.\ Rev.\ Lett.\  {\bf 85} (2000) 2430.

\bibitem{Klinkhamer}F.R.~Klinkhammer and N.S.~Manton,
                    Phys. Rev. {\bf D30} (1984) 2212.

\bibitem{Sstar} F.R.~Klinkhamer, 
                Nucl. Phys. {\bf B 410} (1993) 343;
                Phys. Lett. {\bf B246} (1990) 131. 

\bibitem{BrihayeKunz}Y.~Brihaye and J.~Kunz,  
                     Phys. Rev. {\bf D50} (1994) 4175.

\end{thebibliography}
\end{document}